\documentclass[fleqn,10pt]{wlscirep}
\usepackage[utf8]{inputenc}
\usepackage[T1]{fontenc}
\usepackage{caption}
\usepackage{graphicx}
\usepackage{xcolor}
\usepackage{colortbl}
\usepackage{tabularx}
\newcolumntype{C}[1]{>{\centering\arraybackslash}p{#1}}

\title{Measuring social mobility in temporal networks}

\author[1,*]{Matthew Russell Barnes}
\author[2]{Vincenzo Nicosia}
\author[1]{Richard G. Clegg}
\affil[1]{School of Electronic Engineering and Computer Science, Queen Mary University of London, UK}
\affil[2]{School of Mathematical Sciences, Queen Mary University of London, UK}
\affil[]{matthew.barnes@qmul.ac.uk}


\begin{abstract}
In complex networks, the ``rich-get-richer'' effect (nodes with high degree at one point in time gain more degree in their future) is commonly observed. In practice this is often studied on a static network snapshot, for example, a {\it preferential attachment} model assumed to explain the more highly connected nodes or a {\it rich-club} effect that analyses the most highly connected nodes. In this paper, we consider temporal measures of how success (measured here as node degree) propagates across time. By analogy with {\it social mobility} (a measure people moving within a social hierarchy through their life) we define hierarchical {\it mobility} to measure how a node's propensity to gain degree changes over time. We introduce an associated taxonomy of temporal correlation statistics including {\it mobility}, {\it philanthropy} and {\it community}. Mobility measures the extent to which a node's degree gain in one time period predicts its degree gain in the next. Philanthropy and community measure similar properties related to node neighbourhood. 

We apply these statistics both to artificial models and to 26 real temporal networks. 
We find that most of our networks show a tendency for individual nodes and their neighbourhoods to remain in similar hierarchical positions over time, while most networks show low correlative effects between individuals and their neighbourhoods.
Moreover, we show that the mobility taxonomy can discriminate between networks from different fields.
We also generate artificial network models to gain intuition about the behaviour and expected range of the statistics. 
The artificial models show that the opposite of the ``rich-get-richer'' effect requires the existence of inequality of degree in a network. 
Overall, we show that measuring the hierarchical mobility of a temporal network is an invaluable resource for discovering its underlying structural dynamics.
\end{abstract}
\begin{document}

\flushbottom
\maketitle

\section{Introduction}
\label{sec:introduction}

In sociology and economics an important characteristic is social mobility: roughly speaking the ability of an individual to improve (or worsen) their position within a social hierarchy throughout their lifetime. An analogous situation within a temporal network might be to ask whether the number of connections a node makes in one time period correlates highly with the number of connections the same node makes in the next time period. In models such as the widely used {\it preferential attachment} model~\cite{Barabasi1999EmergenceNetworks} ``wealthy" nodes continue to accrue connections (probability of a node gaining a new link is proportional to its degree) and nodes with few connections are unlikely to overtake them. Other models such as ``fit-get-richer"~\cite{Fortunato2006Scale-freeRanking} have a similar effect, the probability of a node gaining a link is proportional to some fitness parameter for the node and nodes with lower fitness will always gain links more slowly. On the other hand models with a ``hot-get-richer" property~\cite{Fire2020TheTime} allow nodes to have a temporary burst of popularity that may allow new nodes to overhaul older ones in attracting attention. However, in the temporal networks literature there is no good measure of the strength of this effect~\cite{holme2012temporal,masuda2016guide}

In this paper, we introduce {\it hierarchical mobility}, which measures the propensity of nodes that gain links to continue to do so. We define the statistic {\it mobility} for complex networks, a network-wide measure of the tendency of nodes which gain links to continue to gain links. Specifically, mobility is the Pearson correlation between the number of links a node attracts in one time period and the number of links a node gains in a subsequent time period. If the mobility correlation is high (near one) then rich nodes remain rich and poor nodes remain poor (this is also known as the Matthew effect~\cite{perc2014matthew}). Conversely, a zero or even a negative mobility means a more fluid hierarchy where individual nodes change their position more readily and a large number of links gained in one time period does not imply a large number of links gained in a subsequent time period.

We expand on this idea by taking into account the influence between individual nodes and their neighbourhood. This allows us to introduce our mobility taxonomy of statistics which measure the effects between individuals and their neighbourhoods. The taxonomy includes the well-known {\it assortativity} but also new statistics such as {\it neighbour mobility}, the logical extension of {\it mobility} applied to neighbourhoods, {\it philanthropy} and {\it community}. The philanthropy statistic measures whether the neighbours of rich individuals benefit from the rich individual's success or, in network terms, if there is a correlation between individual node degrees and the gain in degree of their neighbours at some later time. The community statistic, in contrast, measures whether rich neighbours create a benefit for an individual or, in network terms, does the average degree of an individual nodes' neighbours correlate to a later gain in degree for that node. 

To investigate the statistics firstly we use artificial models that attempt to maximise or minimise the statistics within the taxonomy to understand the viable range for all statistics. Following this we test the models on a corpus we collected of twenty six temporal networks. The networks are organised into field (social network, citation network etc). Finally, we use PCA to see how the types of network are separated by these taxonomy measures.

\section{Methods}
\label{sec:method}

\subsection{Temporal Networks}
Our analysis is grounded in temporal graph analysis techniques, which model how interactions (edges) between entities (nodes) evolve over time. A temporal graph $G$ is defined as existing from time $0$ until $\infty$, with nodes $V$ and temporal edges $E$. It is built out of edge events $(n,m,t)$ where $n$ and $m$ are nodes which are connected by an edge at time $t$. In this case, the edges are undirected and are defined by a pair of nodes $e=(n,m) \in E$, where the ordering of the nodes is not important, and $n,m \in V$ but $n\neq m$. If we let $0\leq t_1\leq t_2\leq\ldots\leq t_n< \infty$, where $t_n$ is the number of events, then the edge events make up a set
\begin{equation}
T = \{(n_i, m_i, t_i): i=1,2,\ldots\}
\end{equation}
where the same edge $(n, m)$ can exist in many events, and at multiple times. This set does not need to be ordered by index, as what is important is the times they occur.

The events can be aggregated by time, i.e. all of the events which occur at time $t$, into a graph $G(t) = (V(t),E(t))$ containing only the edges which appear in the appropriate events. This is not necessarily a ``complex" graph, as it could be the case that only one edge exists at a particular $t$. These snapshot graphs can be used to represent the temporal graph
$G(0,\infty) = \{G(t_1),G(t_2),\ldots)\}$
and in a similar way, we define the graph $G(q,r) = (V(q,r),E(q,r))$ that exists in ``time window'' $(q,r)$, from time $q$ to $r$, where $0<q<r<\infty$, as the aggregation of all edge events $(n_i, m_i, t_i)$ where $q\leq t_i < r$. Excluding the timestamp $r$ removes double counting of adjoining time windows.

\subsection{Mobility Taxonomy}
\label{subsec:mob_tax}

Consider two time periods, one running from $q$ to $r$ and the next running from $r$ to $s$. We are interested in how nodes and neighbourhoods evolve in these two time periods. Let $K(q,r)$ and $K(r,s)$ be the set of degrees of all nodes in these time periods and let $L(q,r)$ and $L(r,s)$ be the set of mean degrees of neighbours. These quantities are defined rigorously in the next section. To measure the association of the degree hierarchy level of nodes, and their neighbours, with their hierarchy level in a subsequent time period we introduce the following mobility taxonomy where the Pearson correlation is measured between all possible combinations of $K$ and $L$. There are six combinations, see table~\ref{tab:mob_tax} that correlate node degree and average neighbourhood degree in time windows one and two. These are {\it mobility}, {\it neighbour mobility}, {\it philanthropy}, {\it community}, {\it assortativity} and {\it consistent assortativity}. Each represents correlating the degree and/or average neighbourhood degree of all nodes in a network that exist between two adjoining time windows. Assortativity is well-studied in network science and consistent assortativity is just a variant of this. These two statistics have no real time aspect so we drop them from further discussion.

In summary the four major statistics we will consider in this taxonomy are: mobility which measures the correlation between the degree a node gains in time period one and the degree it gains in time period two; neighbourhood mobility, which is the same operating on neighbourhoods not individual nodes; community, which measures the correlation between a node's degree in time period two and its neighbourhood's degree in time period one; and philanthropy, which measures the correlation between a node's degree in time period one and its neighbourhood's degree in time period two. Community can be thought of as measuring whether having rich neighbours helps you become rich at a later time. Philanthropy can be thought of as measuring whether a rich node spreads their success to their neighbourhood at a later time.

\vspace{1em}
\begin{minipage}{0.9\textwidth}
\begin{tabular}{| C{2.7cm} | C{2.7cm} | C{2.7cm} | C{2.7cm}|} \hline
\cellcolor{yellow} Correlation & \cellcolor{yellow} Degree (period 2) $K(r,s)$ & \cellcolor{yellow} Mean neighbour degree (period 1) $L(q,r)$ & \cellcolor{yellow} Mean neighbour degree (period 2) $L(r,s)$ \\ \hline
\cellcolor{yellow} Degree (period 1) $K(q,r)$ \vskip 0cm & Mobility & Assortativity & Philanthropy \\ \hline
\cellcolor{yellow} Degree (period 2) $K(r,s)$ \vskip 0cm & \cellcolor{lightgray} & Community & Consistent assortativity \\ \hline
\cellcolor{yellow} Mean neighbour degree (period 1) $L(q,r)$ & \cellcolor{lightgray} &  \cellcolor{lightgray} & Neighbour mobility \\ \hline
\end{tabular}
\captionof{table}{The taxonomy of mobility related aspects for two time periods, the first running from $q$ to $r$ and the second from $r$ to $s$. The columns correspond to X in the Pearson correlation (equation (\ref{eq:pearson})) and the rows to Y. The cells represent the name we give to the correlation statistic from the variables in the row and column.}
\label{tab:mob_tax}
\end{minipage}
\vspace{1em}

\subsubsection{Mathematics of the Mobility Taxonomy}

This section gives the exact details of how the mobility taxonomy is calculated. The correlations are between node degree (or mean neighbour degree) in two time periods $(q,r)$ and $(r,s)$. The sets to be correlated must contain measured values corresponding to a consistent node set between the two windows, this way we are correlating consistent graphs over time. Therefore, we define the set of consistent nodes as $V_c \equiv  \{n_1,n_2,\ldots\} =  V(q,r) \equiv \{n_1(q,r),n_2(q,r),\ldots\} $, as this is the first time window we are correlating with. The number of edges that node $n$ in $V_c$ takes part in during the time window $(q,r)$ is the degree of that node 
\begin{equation}
\label{eq:degree}
k_n(q,r) = \left\{ 
  \begin{array}{ c l }
    \sum_{m\in V(q,r)} a_{mn}(q,r) & \quad \textrm{if } n \in G(q,r) \\
    0                 & \quad \textrm{otherwise}
  \end{array}
\right.
\end{equation}
where $n\in V_c$ and $a_{mn}(q,r)$ is the element in the adjacency matrix equivalent of $G(q,r)$ corresponding to an edge between nodes $m$ and $n$. The conditional statement accounts for nodes in $V_c$ which do not appear in $G(q,r)$ --- this is crucial for time windows other than $(q,r)$. Performing this for all of the nodes in $V_c$ we produce the tuple $K(q,r) = (k_n(q,r): n\in V_c)$.

The {\it mobility} of a temporal network is the Pearson correlation of the degree of nodes appearing in $G(q,r)$ with the degree of the same nodes as they appear in $G(r,s)$ where $q,r$ and $s$ are times that satisfy $0<q<r<s<\infty$. The Pearson correlation is defined as 
\begin{equation}
\label{eq:pearson}
\rho_{(X,Y)} = \frac{E[(X-\mu_X)(Y-\mu_Y)]}{\sigma_X \sigma_Y}
\end{equation}
where $X$ and $Y$ are tuples of equal length. For mobility we substitute in $X = K(q,r)$ and $Y = K(r,s)$, which is the correlation of the degrees of the consistent node set $V_c$ in both time windows $(q,r)$ and $(r,s)$.

To calculate community the average neighbourhood degree needs to be defined. First, we define the consistent set of the neighbours of nodes $n \in V_c$ as $N_c \equiv \{N_{n_1},N_{n_2},\ldots\} = N(q,r) \equiv \{N_{n_1}(q,r),N_{n_2}(q,r),\ldots\}$, where $N_n(q,r) = \{m\in V_c: m,n\in E(q,r)\}$ which denotes all nodes $m$ which exist in the edges where node $n$ also exists during time window $(q,r)$. From this, the average degree of the nearest neighbours of node $n$ in $V_c$ is defined
\begin{equation}
\label{eq:and}
l_n(q,r) = \frac{1}{|N_n|}\sum_{m\in N_n} k_m(q,r)
\end{equation}
where $N_n \in N_c$ and $k_m(q,r)$ is defined in equation (\ref{eq:degree}). This is gathered into the tuple of all nodes is $L(q,r) = (l_n(q,r): n\in V_c)$. Therefore, to calculate community we substitute $X = L(q,r)$ and $Y = K(r,s)$ into equation (\ref{eq:pearson}). Also, for philanthropy we simply reverse these time windows and substitute $X = K(q,r)$ and $Y = L(r,s)$. Note that even a node that does not exist in time window $(r,s)$ will calculate its average neighbour degree using its consistent neighbourhood as defined in $N_c$.

Now we have the tools to calculate all of the rest of the aspects of the mobility taxonomy. For neighbour mobility, we substitute $X = L(q,r)$ and $Y = L(r,s)$ into equation (\ref{eq:pearson}), assortativity uses the substitutions $X = K(q,r)$ and $Y= L(q,r)$ and finally, consistent assortativity uses the substitutions $X = K(r,s)$ and $Y= L(r,s)$. For clarity all of the substitutions for each mobility taxonomy aspect are outlined in Table~\ref{tab:mob_tax}. In practice, the temporal graph analysis tool Raphtory~\cite{steer2020raphtory} is used to calculate the raw degree and average neighbour degree numbers before performing these correlations. Then these raw numbers are correlated together as outlined above and plotted. 

\section{Results}

\label{sec:results}

We experiment first on artificial models to establish a baseline of how the taxonomy behaves using artificial network models. This enables us to understand the effective range of each statistic in the taxonomy. While, theoretically, a correlation is in the range $[-1,1]$ in practice they cannot achieve the extremes of this range. Following this we use a corpus of twenty six real-world networks grouped by collection type (social network, communication network etc) and link creation type (whether the network grows by adding individual links, cliques or other structures). 

\subsection{Artificial Models}
\label{subsec:models}

First we establish an understanding of the behaviour of each of the four statistics in the taxonomy using a variety of artificial models. For example, in particular our interest is to establish the effective working range for each by attempting to maximise or minimise that statistic. We do this using twelve candidate artificial models, the details of which are given in Table S2. The models are chosen from the literature and from insights about what behaviour might be likely to maximise or minimise statistics within the taxonomy. Each model grows a network following the structure of the classic preferential attachment model of~\cite{Barabasi1999EmergenceNetworks} model: add one new node and link it to three existing nodes by some rule. In preferential attachment nodes are picked with a probability proportional to their degree. For each experiment we choose a statistic to minimise or maximise. We first use preferential attachment to form a seed network of 3,000 nodes (9,000 links). We then follow this procedure for our target statistic $S$ that we wish to maximise/minimise:
\begin{enumerate}
    \item For each of our twelve models create twelve candidate networks by adding another 1,000 nodes (3,000 links) -- we call this one time slice.
    \item Measure the statistic $S$ for each of the twelve candidates by comparing the ``current" time slice with the previous.
    \item Choose the candidate which best maximised (or minimised) $S$ and take that network as the canonical continuation of the artifical network.
    \item If we have less than ten new time slices go to step 1.
\end{enumerate}

\begin{figure}[!tb]
\centering{\includegraphics[width=0.8\textwidth]{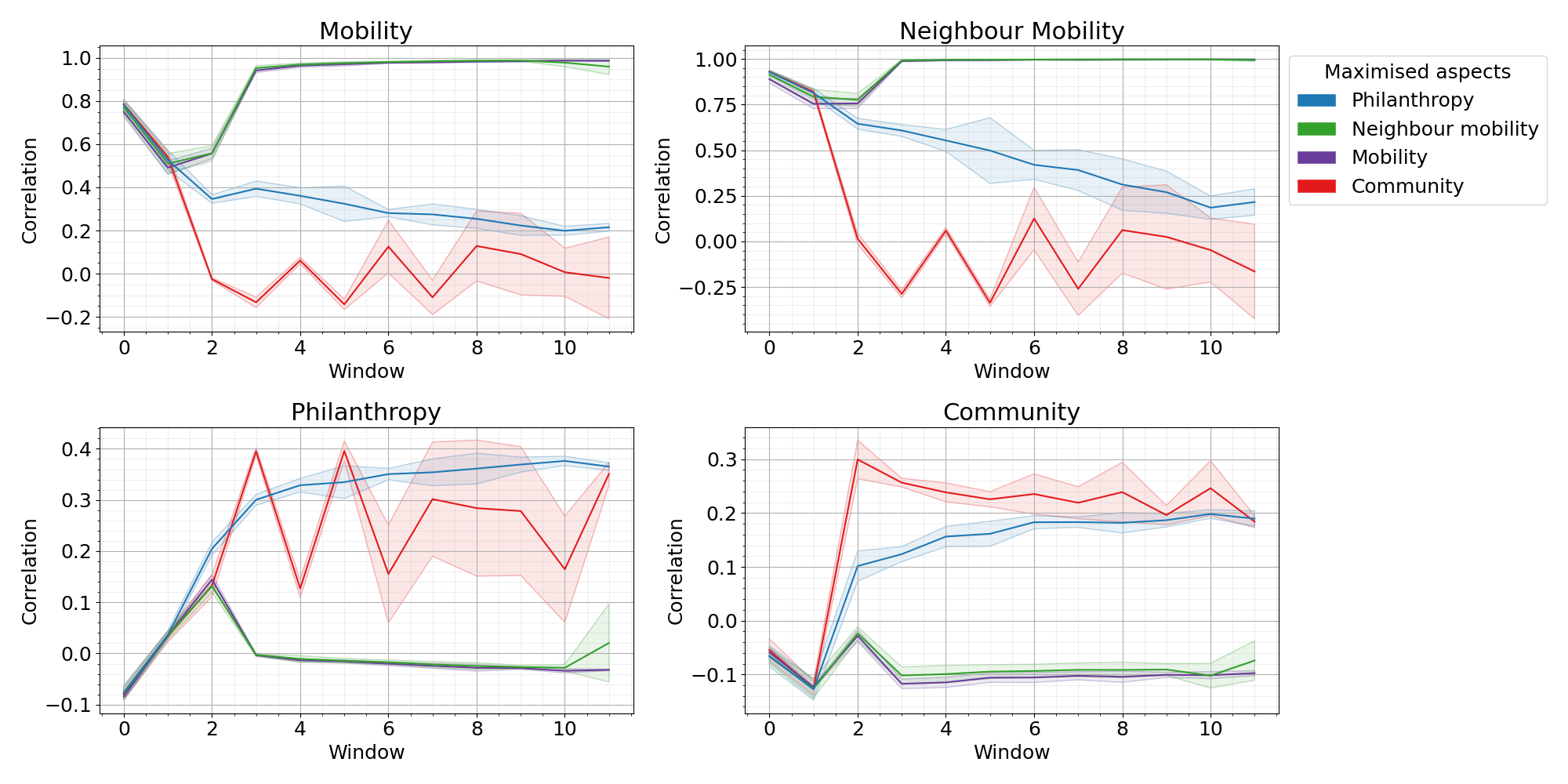}
  \caption{Maximisation of mobility taxonomy aspects for ten time windows. Each graph shows one statistic in the taxonomy and the colour represents which statistic is being maximised. The shading is the standard deviation over ten separate realisations.}
\label{fig:optimiser_maximised}
}
\end{figure}

\begin{figure}[!tb]
\centering{\includegraphics[width=0.8\textwidth]{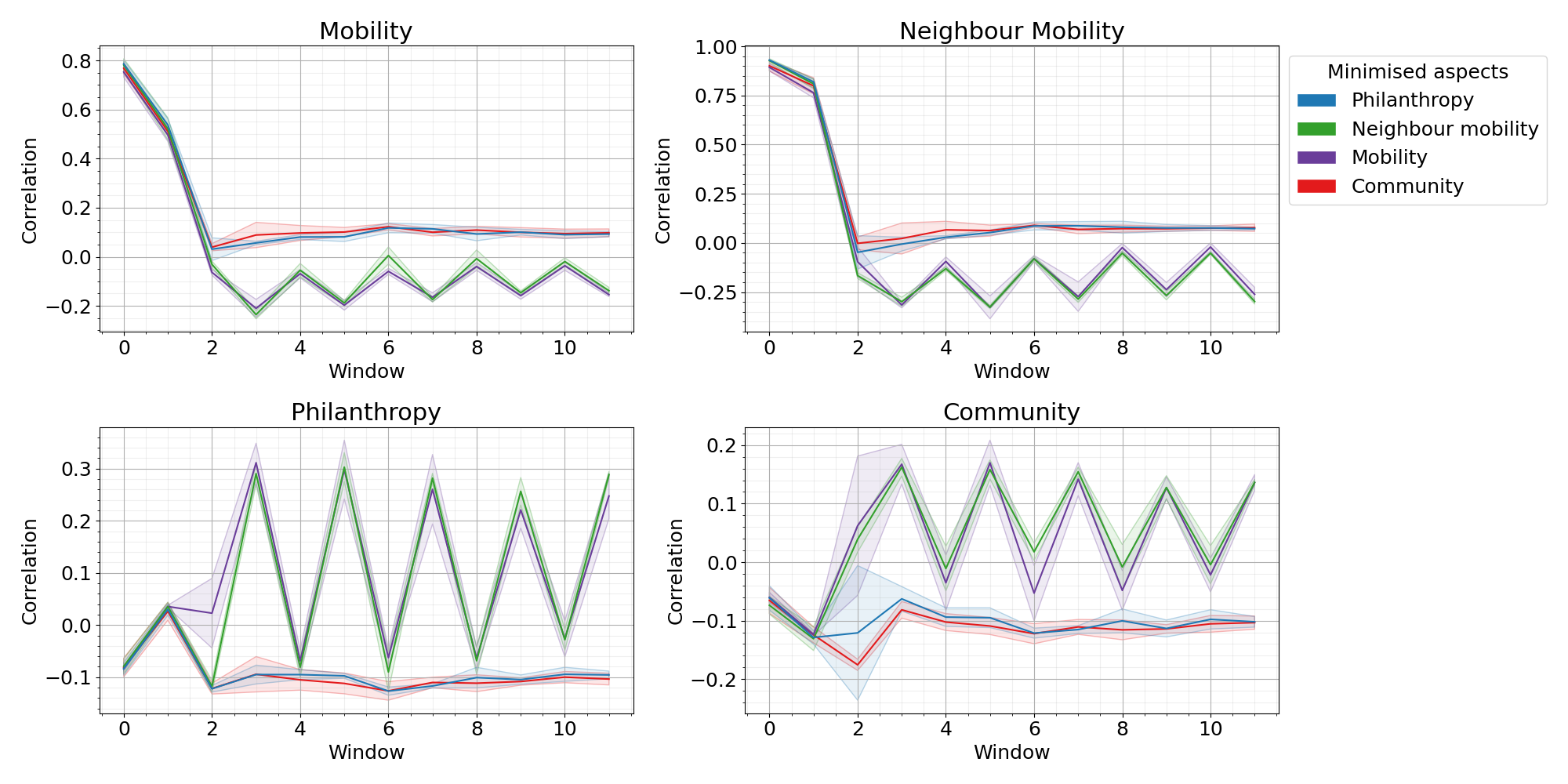}
  \caption{Minimisation of mobility taxonomy aspects for ten time windows. Each graph shows one statistic in the taxonomy and the colour represents which statistic is being minimised. The shading is the standard deviation over ten separate realisations.}
\label{fig:optimiser_minimised}
}
\end{figure}

Figure~\ref{fig:optimiser_maximised} shows each statistic for the four different maximisation experiments and figure~\ref{fig:optimiser_minimised} the equivalent for the minimisation experiments. Mobility and neighbour mobility (the upper graphs in each figure) often have very similar behaviour. Both can reach the maximal correlation of 1.0. Maximising one maximises the other and similarly minimising one minimises the other. This is intuitive: if all nodes have a very high/low correlation in the degree gained between timeslice two and timeslice two both mobility and neighbourhood mobility will be high. Minimising these aspects gets values of -0.2 and -0.25 respectively. Philanthropy and community (the lower graphs in each figure) also have similar behaviour in some experiments. Neither gets very high compared with the other two statistics with philanthropy reaching a maximum of 0.4 and community a maximum of 0.3. For minimisation philanthropy can be as low as -0.1 and community -0.15 (but this is a single point). Again it makes sense the community and philanthropy operate in a more limited range of values. It is very easy to think of a strategy where every node gains many links in time period two if and only if it gains many links in time period one, such as preferential attachment. For community (and similarly 
philanthropy), where the a node will only gain links in period two if and only if their neighbourhood had high average in period one, it is much harder to think of such a strategy. 

It is noticeable that in some experiments the statistics rise and fall by large amounts every iteration. This happens for community maximisation (the red lines in figure~\ref{fig:optimiser_maximised}) and for mobility and neighbour mobility minimisation (the green and purple lines in figure~\ref{fig:optimiser_minimised}). This is because those maximisation/minimisation strategies actually alternate the model they use between different time slices. The models used in each time slice for each experiment are included in supplemental material. This happens because strategies to produce negative mobility/neighbour mobility correlations changes. If a model too rigorously minimises by preferentially connecting to low degree nodes then this same strategy will not work in the next time period. 

Table~\ref{tab:artificial_ranges} shows the ranges for the four statistics as inferred from the artificial models. Of course these tests were limited in that only a certain class of models were tested and real data may have results outside these limits but these values give an intuition that, for example, 0.3 is a very high value for philanthropy or community but only a medium value for mobility. 

\begin{table}
    \centering
    \begin{tabular}{|c|c|l|} \hline
       Aspect  & Effective Range & Comment \\ \hline
       Mobility  & (-0.2,1) & Alternating strategies minimises this. \\
       Neighbour mobility & (-0.25,1) & Alternating strategies minimises this. \\
       Philanthropy & (-0.1,0.4) & Alternating strategies maximise this. \\
       Community & (-0.15,0.3) & Alternating strategies maximise this. \\ \hline
    \end{tabular}
    \caption{Effective ranges of the statistics within the taxonomy.}
    \label{tab:artificial_ranges}
\end{table}

\subsection{Data Corpus}
Our corpus of 26 network datasets is taken from real world systems and have a collection type (Social, Citation, Economic, Co-occurrence, Computer,  Contact and Transport) based on the kind of data collected, and a link creation type (Star, Bipartite, Individual, Clique and Spatial) based on how the network grows by adding nodes and links: star networks grow by adding a node and some links from it (for example a citation network); bipartite networks have two node types and links are between those types (for example customers and products they buy); a clique network grows by adding cliques (in a co-authorship network each paper will add a clique for all authors); spatial networks are those where links are defined by nodes being ``close" in space at a particular time (for example a contact network); finally networks which don't fall into this scheme are designated individual. Further detail about the corpus and its types is given in Section S1 and Table S1.

\begin{figure}[!tbp]
\centering{\includegraphics[width=0.8\textwidth]{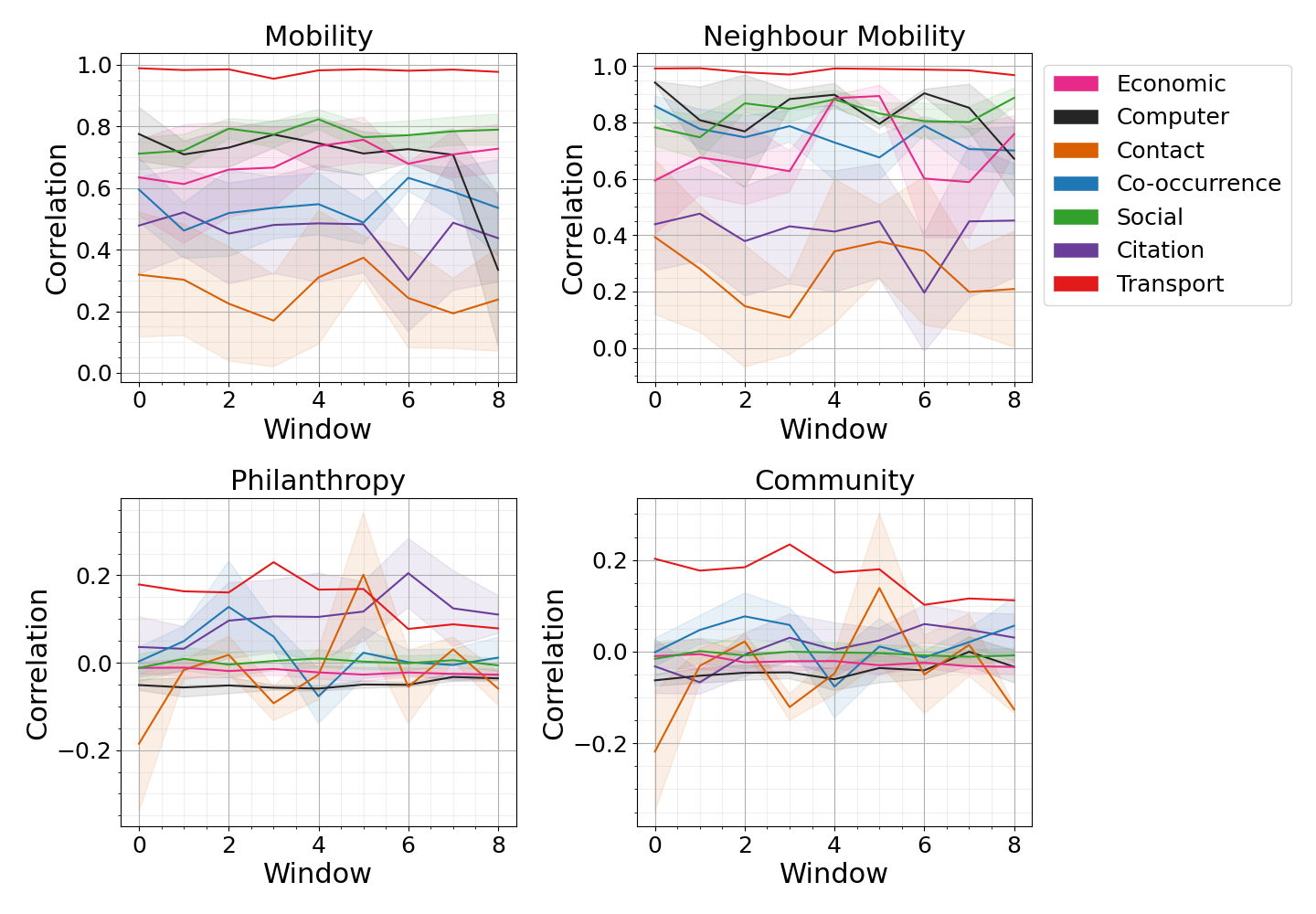}
  \caption{All networks' mobility taxonomy correlations in the corpus, with the shaded area representing one standard error of the mean and coloured by collection type. The x-axis represents the ordered window numbers starting at the first timestamp of the network. The y-axis shows appropriate statistics.}
\label{fig:taxonomy_over_time}
}
\end{figure}

In Figure~\ref{fig:taxonomy_over_time}, we plot each of the mobility taxonomy aspects for all networks in the data corpus as outlined in Table S1. Each collection type is given a colour and shading which shows the mean and one standard error either side. Each individual data set is plotted in supplemental material Figure~S1. Comparing between our measures we can see that, as with the artificial data mobility and neighbour mobility are closely correlated with each other and, similarly, philanthropy and community show similar behaviour. We can see that the values are within the ranged predicted using the artificial model studies with the sole exception of the window zero for philanthropy and community which is slightly lower (looking at the individual results this is result of a single data set, set q, with a very low value at this point). 

For the mobility and neighbourhood mobility measures the single transport network and the social networks show the highest values. This is not unexpected, these are networks where having existing connections is a large benefit (the transport network is flights between airports) and a good predictor of connection numbers in subsequent time periods. Conversely contact networks and spatial networks showed a much lower effect. Citation networks have been associated with a ``hot-get-richer" effect~\cite{Fire2020TheTime} which would lower the time correlation (a paper which is ``hot" and gains citations in one time period finds it hard to sustain this in subsequent time periods while new ``hot" papers arrive).

For philanthropy and community, as predicted by the artificial model results, the available range is smaller. The transport network and the citation network have high philanthropy results. In the case of an airport network this can be translated as `connecting to an airport with many connections brings more connections in the future" in the case of a citation network this can be translated as `citing a well-cited paper brings citations", a result which might arise from sub-disciplines in academia changing their popularity overall spurred by a single ``breakthrough" that becomes popular. Computer networks show consistently negative philanthropy and community. These networks often show disassortativity, high degree nodes link preferentially to low degree nodes, which could explain this finding.

\subsubsection{Principal Component Analysis}
\begin{figure}[!tbp]
\centering{
  \includegraphics[width=0.7\textwidth]{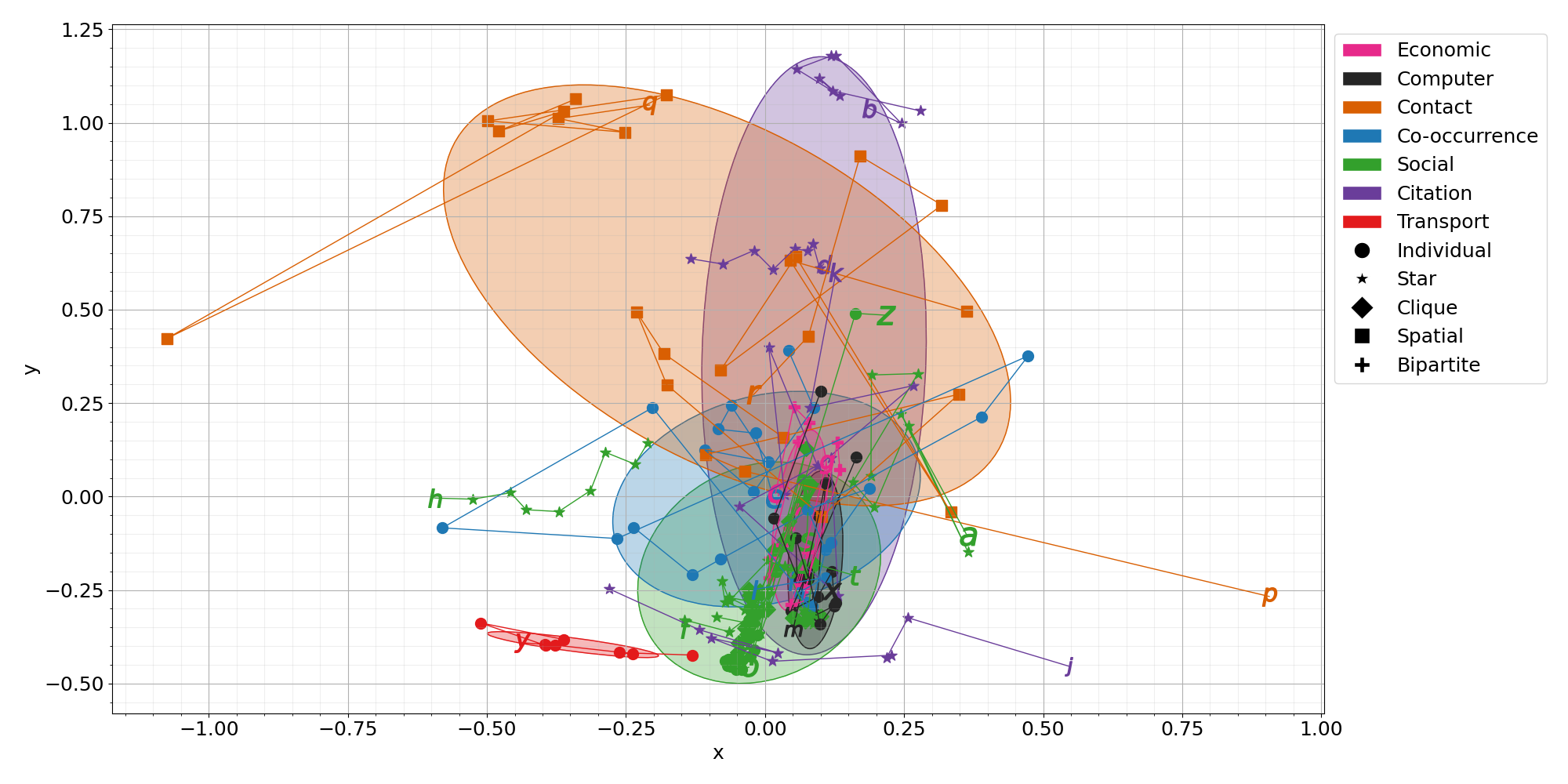}
  \caption{Principal component analysis of the mobility taxonomy with Gaussian Mixture Model ellipses. The plotted lines denote the datasets through their time-steps, starting with the letter from Table S1, each step plotted with a shape corresponding to its link creation type and the colour corresponding to its collection. The x component takes its largest contributions from the negative of {\it philanthropy} and {\it community}, whereas the y component is mostly the negative of {\it mobility} and {\it neighbour mobility}.}
  \label{fig:PCA}
}
\end{figure}

To visualise how the taxonomy separates our types of data we perform a Principal Component Analysis (PCA) using the six mobility taxonomy aspects as the dimensions of the decomposition. This is performed in the first window of each network, and this consistent covariance matrix is applied to all subsequent windows. In Figure~\ref{fig:PCA}, we plot the first two PCA components (the two highest variance components of the consistent covariance matrix) for every network with one point for every time window studied. We used a PCA over non-linear alternatives to ensure the axes remained easily interpretable. These components are a combination of the negative of philanthropy and community (with some {\it assortativity} and {\it consistent assortativity}) on the x axis and the negative of mobility and neighbour mobility on the y axis, see Table~\ref{tab:PCA_com} for exact numbers. 

\begin{table}[!tbp]
    \resizebox{\textwidth}{!}{\begin{tabular}{c|c|c|c|c|c|c}
        Axis & Mobility & Neighbour mobility & Philanthropy & Community & Assortativity & Consistent assortativity  \\
         \hline \hline
        x & $-0.13$ & $-0.15$ & $-0.52$ & $-0.53$ & $-0.45$ & $-0.46$ \\
        \hline
        y & $-0.69$ & $-0.69$ & $0.13$ & $0.18$ & $0.07$ & $-0.01$ \\
        \hline
        
    \end{tabular}}
        \caption{Components for each axis of the PCA plot, Figure~\ref{fig:PCA}.}
    \label{tab:PCA_com}
\end{table}

Each window is connected via a straight line to each subsequent window to track the time evolution of each network. The collection types from Table S1 of each network are represented by the colour of the markers and lines and link creation types by the marker shapes. Finally, each collection type group of networks is processed using a Gaussian Mixture Model with the resulting ellipse plotted to represent the general spread of each group.

From Figure~\ref{fig:PCA} we can see that the social networks (eleven networks) are confined to a remarkably small area of the statespace. They have nearly no overlap with, for example, transport and contact networks. Looking at the ovals for the network types it is clear that there is considerable separation within the statespace introduced by the network types, although it is not perfect, and we would not expect it to be. Though there are exceptions, most networks occupy a relatively small space within the network in their lifetimes (we appreciate this is not easy to discern in the centre of the diagram) indicating that many individual networks in the corpus can be characterised by these six statistics. There are some datasets (notably p and q) where a single point is very different to the others (because that data set has an event at that time) and others (the co-occurrence network c) that take up a larger area in the state space. 

\subsection{Equality measurements}

While not entirely related it is useful to look at the networks through the lens of equality. Here we use the Gini coefficient~\cite{Gini1912VariabilitaMutabilita} as a measure since it is widely used in economics.
It is usually used as a measure of income inequality and is given by
$$
    G = \frac{\sum^{n}_{i=1}\sum^{n}_{j=1}|x_{i}-x_{j}|}{2n^2\Bar{x}}
$$
where $x_{i}$ is the income of person $i$, $n$ is the number of persons, and $\Bar{x}$ is the mean income of the population. Here we simply take $x_{i}$ the income of person $i$ as $d_i$ the degree of node $i$ and use Gini as a measure of the degree inequality of the node distribution. A Gini coefficient of 1 indicates perfect inequality and 0 indicated perfect equality. 

\begin{figure}[!tbp]
\centering{
  \includegraphics[width=0.7\textwidth]{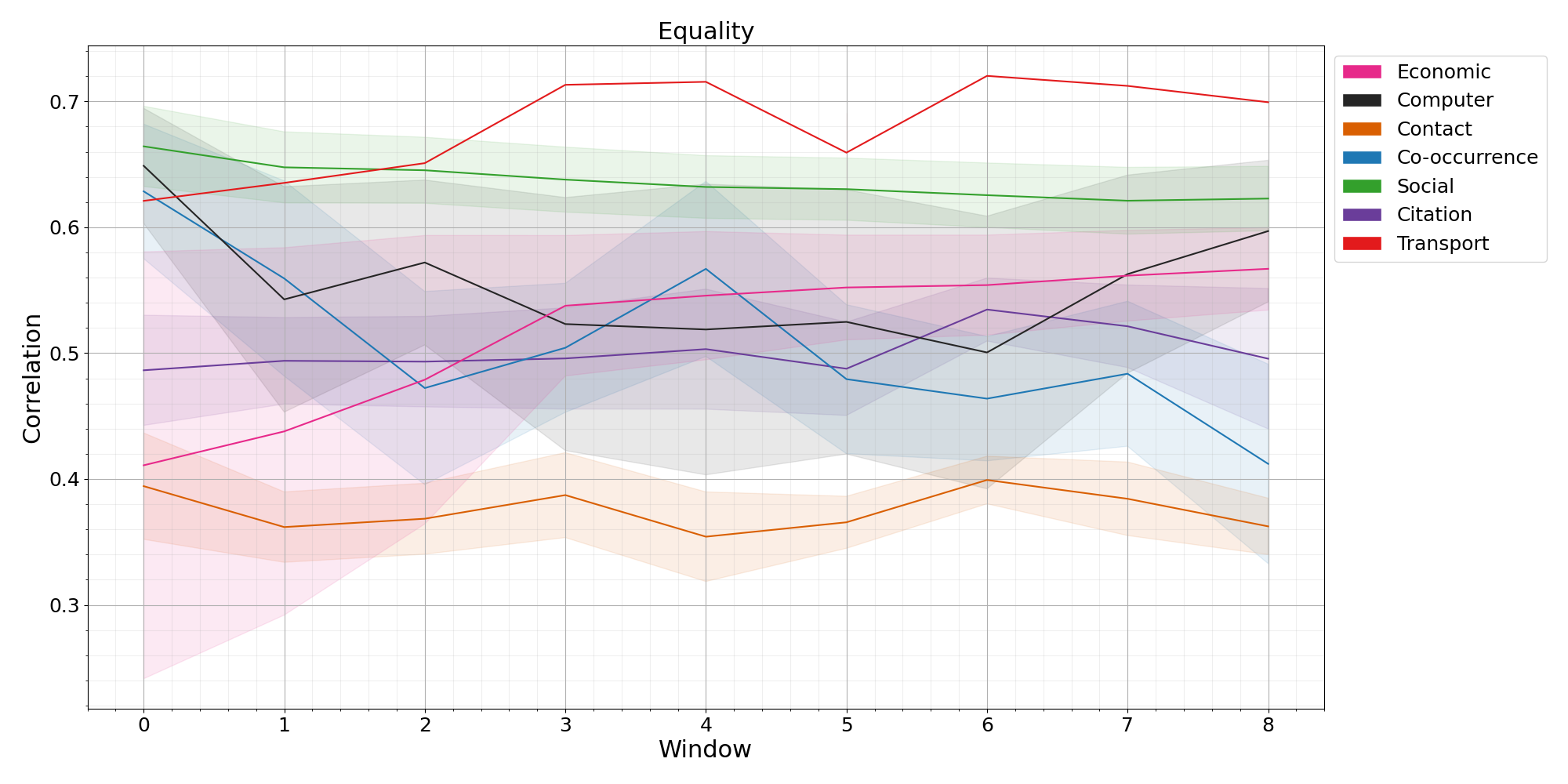}
  \caption{Gini coefficient for the datasets in our corpus. The mean is taken over each of our types and the standard error of the mean indicated}
  \label{fig:gini}
}
\end{figure}

Figure~\ref{fig:gini} shows the result on our corpus. We can see considerable variety with the transport and social networks being the most unequal (higher Gini). The contact networks are the most equal which makes sense as those networks often have a spatial constraint with two consequences: it is hard for one node to be near many without all those neighbour nodes having a similarly high degree (they are part of the large group). 

\section{Related Work}
\label{sec:related_work}
Tracking the hierarchical mobility of human societies between generations has been a focus of Sociologists and Economists for decades. Traditionally, the way to model a hierarchy is by ranking individual things by some metric. Sociologists tend to focus on occupational ``class" hierarchies~\cite{Szreter1984TheOccupations} which are qualitative and subjectively ranked~\cite{Erikson1992TheFlux}. One sociological measure of mobility is the ``Log-Multiplicative Layer Effect Model''~\cite{Breen2004SocialEurope} which compares two matrices (called ``layers'' or ``generations'') of class associations by assuming a uniform multiplicative association. This uniform association removes much of the much needed nuance between inter-generational class associations. 

Economists tend to focus on financial income ``bands"~\cite{Mayer2005HasChanged} which are quantifiable and inherently ranked. A widely used mobility measurement used in economics is the Pearson correlation coefficient~\cite{Solon2001IntergenerationalMobility}
$\beta = r(\ln{Y_c},\ln{Y_p})$
where $r(X,Y)$ is the Pearson correlation between numerical series $X$ and $Y$. This is usually used in conjunction with
$\ln{Y_c} = \alpha + \beta_p\ln{Y_p} + \epsilon_c$
where $Y_c$ is the income of children, and $Y_p$ is the income of parents, $\alpha$ is a constant and $\epsilon_c$ is a fitted constant. 

A more direct approach to analysing hierarchical dynamics was taken in a recent paper~\cite{iniguez2022dynamics} by observing the dynamics of a variety of rank-ordered lists. The authors found that systems where the top 100 ranks are more ``open" to new inhabitants experience higher hierarchical mobility, as opposed to less open systems where mobility is low. Also, the rest of the list is invariably much less stable than these top 100 ranks.

Temporal networks model the time evolution of pair-wise interactions that are inherent in natural networks. The analysis of such networks is becoming more common~\cite{masuda2016guide} as the traditional approach of analysing a single static snapshot of networks does not allow for the analysis of network dynamics. Instead, comparing multiple snapshots of a network taken within subsequent time windows allows for us to track the time evolution of important nodes~\cite{Fire2020TheTime,Nsour2020Hot-Get-RicherModel} and their association with the nodes they connect to~\cite{Zhou2020UniversalNetworks,pedreschi2022temporal}. In this paper we expand on these two ideas by using similar metrics and correlating them between time windows. We compare the outcomes of these correlations across many artificial models and real world networks.

Building artificial networks with particular characteristics using models which apply particular growth rules over time to nodes and edges is a very well trodden field. Of course, there are the two models referenced in the introduction, the Barab\'{a}si-Albert (BA)~\cite{Barabasi1999EmergenceNetworks} model and the Fortunato~\cite{Fortunato2006Scale-freeRanking} model which both replicate power-law degree distributions using simple rules. However, complex models are becoming more prevalent, such as models which mimic assortativity in social networks~\cite{Zhou2020UniversalNetworks}.

In this paper, we build on the economic hierarchical mobility measurements by removing the need for classes or bands by increasing the resolution to individual people (or nodes). We also incorporate the interactions between people by modelling them as temporal networks and we increase the resolution of the time windows from generations to any ``useful" time-frame. Finally, we use Pearson correlations to measure the time evolution of a network's ``degree hierarchy". The mechanisms for time evolution involve timestamped node and edge additions, and windows of time for which the network is aggregated and analysed in comparison to other time window graphs.

We briefly look at equality in our data through the lens of the Gini coefficient in our results section. Some authors have considered how inequality changes over time by conducting either simulated or real life experiments on networks. In~\cite{cui2023kinetic} the authors create an artificial trading model to study the effect on wealth inequality which demonstrates that the Matthew effect can arise from relatively few experimental assumptions. 
In~\cite{nishi2015inequality} the authors use an experimental economics approach to look at how people redistribute wealth in a social network, reporting initially high inequalities dropping and initially low inequalities rising. It is hard to generalise from their findings to the current work.

\section{Conclusions}
\label{sec:conclusion}
In this paper, we introduce the mobility taxonomy which correlates individual nodes and shows how they evolve over time. We split networks into time slices and investigate correlations between time slices. Together the statistics within the taxonomy measure the tendency for nodes (and neighbourhoods of nodes) to maintain a growth trajectory over time. The taxonomy has six aspects, here we focus on the four statistics which capture time aspects: {\it philanthropy}, {\it community}, {\it mobility} and {\it neighbour mobility}. 

Each statistic is a correlation measured across time slices in a network and is in the range $[-1,1]$, however, for a variety of reasons the statistics cannot use the full range. Tests on artificial networks to minimise/maximise these statistics showed that mobility and neighbourhood mobility were in the range $(-0.2,1)$ and $(-0.25,1)$ respectively while philanthropy and community had smaller ranges, $(-0.1,0.4)$ and $(-0.15,0.3)$. Different strategies minimised and maximised the statistics with, in some cases, alternating strategies which switched every time slice being the best found.

Tests were performed on a real data corpus with twenty six networks split into different types. They were investigated using the taxonomy statistics and the standard economic measure of inequality the Gini coefficient. Different types of networks produced different measures in this analysis. Social networks and the transport network, for example, showed high mobility (and high inequality). Citation networks and the transport network again showed high philanthropy and community. Computer networks showed consistently negative philanthropy and community. Using PCA to visualise all six statistics in just two dimensions revealed that most (but not all) networks stay within an area of the state space throughout their lifetime. It also showed that some types of networks were confined within a small area of the state space. 

This paper has introduced a taxonomy of correlations which can be used to measure the effects that individuals and their neighbours have on the dynamics of complex systems. The fact that networks usually have consistent values for the statistics across time (as revealed by the PCA) shows that the measure is genuinely capturing a quality of that data set. While the primary aim of the taxonomy is not simply to separate networks by type it was shown that many of the types of network studied had very different behaviour and could be characterised using these statistics. These statistics reveal useful measures about how networks evolve over time and further study could help understanding of network dynamics. 

\bibliography{data_corpus_2023}

\section{Author Contributions}
M.R.B., R.G.C., V.N. conceived the experiments, M.R.B. conducted the experiments, M.R.B. analysed the results. M.R.B., R.G.C. reviewed the manuscript.

\section{Data Availability}
The datasets used and analysed during the current study available from the corresponding author on reasonable request.

\section{Funding}
The authors wish to acknowledge the support of Moogsoft Ltd. for funding this research.

\section{Competing Interests}
The authors declare no competing interests.

\end{document}

\maketitle
\section{Data Corpus Details}
\label{subsec:data_corpus_intro}
There are many different varieties of systems which can be modelled using temporal networks. Therefore, we have gathered a corpus of 26 network datasets taken from real world systems. The networks vary in size from just over 100 edges to more than one million and all networks are treated as undirected and unweighted with added their links never being removed. All of the networks can be seen in Table~S\ref{tab:datasets} with a description for each. 

We have assigned each dataset a ``collection type" based on where the data was collected. Social data is gathered from a social media site, email list or messaging app; nodes are people and edges are emails, messages or ``follows'' between two people. Co-occurrence data is gathered from books, articles and musical notation; nodes are words or musical notes and the edges are created when two nodes exist next to one another in the source material. Citation data is gathered from literature which cites other literature; nodes are pieces of literature and edges are the citations. Economic data is gathered from places where financial transactions are of interest; nodes are products, buyers and sellers and edges are linking buyers/sellers with their products. Computer data is gathered from computer networks; nodes are computers and edges are interactions between the computers. Transport data is gathered from transportation networks; nodes are geographical places and edges are the transportation links between the places. Finally, contact data is gathered when people interact in close proximity, usually using proximity sensors; nodes are people and edges are created when two people are physically close for a threshold of time. 

We also assign a ``link creation type" based on the dynamics of the networks themselves. Stars consist of all edges connecting to one node, for example when one paper cites others. Individual edges are created when only two nodes interact, for example two characters in a book. Cliques are fully interconnected node clusters, for example everyone who answers a question on StackOverflow is connected together. Bipartite networks consist of two separate node pools which only link with each other, and not within themselves. Finally, spatial networks are a mix of many of the others and so have been separated into their own category. 

\begin{table}[!tb]
    \resizebox{\textwidth}{!}{\begin{tabular}{c|c|c|c|c|c|c}
        & Name & Description & Collection & Link Creation & Nodes & Edges  \\
         \hline \hline
        a & College Messages~\cite{Panzarasa2009PatternsCommunity} & Messages between students on a UC-Irvine message board & Social & Star & 1,899 & 59,835 \\
        \hline
        b & SCOTUS Majority~\cite{Fowler2007NetworkCourt} & Legal citations among majority opinions by SCOTUS & Citation & Star & 34,613 & 202,167 \\
        \hline
        c & Amazon Ratings~\cite{Lim2010DetectingBehaviors} & Amazon user connects to all products they have rated & Economic & Bipartite & 3,376,972 & 5,838,041 \\
        \hline
        d & Citation US Patents~\cite{Hall2001TheTools} & Citation among patents in the United States & Citation & Star & 3,774,362 & 16,516,270 \\
        \hline
        e & Classical Piano~\cite{Park2020NoveltyNetworks} & Transitions of chords in western classical piano music & Co-occurrence & Individual & 144,183 & 585,154 \\
        \hline
        f & Email EU~\cite{Paranjape2017MotifsNetworks} & E-mails between users at an EU research institution & Social & Star & 986 & 332,334 \\
        \hline
        g & Procurement EU~\cite{Wachs2021CorruptionPerspective} & Public EU procurement contracts & Economic & Bipartite & 839,824 & 4,098,711 \\
        \hline
        h & Facebook Wall~\cite{Viswanath2009OnFacebook} & Posts by users on other users' Facebook wall & Social & Star & 46,952 & 876,993 \\
        \hline
        i & Lord Of the Rings~\cite{Clegg2009} & Character co-occurrence in Lord of the Rings Trilogy & Co-occurrence & Individual & 139 & 2,649 \\
        \hline
        j & PhD Exchange~\cite{Taylor2017Eigenvector-basedNetworks} & Exchange of PhD mathematicians between unis in the US & Citation & Star & 230 & 9,584 \\
        \hline
        k & Programming Languages~\cite{Valverde2015PunctuatedLanguages} & Influence relationships among programming languages & Citation & Star & 366 & 762 \\
        \hline
        l & Reuters Terror News~\cite{Corman2006StudyingSystems} & Word co-use in Reuters 9/11 coverage & Co-occurrence & Individual & 13,308 & 148,035 \\
        \hline
        m & Route Views~\cite{Clegg2009} & Route Views internet topology & Computer & Individual & 33,804 & 94,993 \\
        \hline
        n & Reddit Hyperlinks Body~\cite{Kumar2018CommunityWeb} & Subreddit-to-subreddit hyperlinks from body of posts & Social & Clique & 35,776 & 286,561 \\
        \hline
        o & Reddit Hyperlinks Title~\cite{Kumar2018CommunityWeb} & Subreddit-to-subreddit hyperlinks from title of posts & Social & Individual & 54,075 & 571,927 \\
        \hline
        p & Hypertext Conference~\cite{Isella2011WhatsNetworks} & Contacts among attendees of ACM Hypertext 2009 & Contact & Spatial & 113 & 20,818 \\
        \hline
        q & Infectious~\cite{Isella2011WhatsNetworks} & Contacts during Infectious SocioPatterns 2011 event  & Contact & Spatial & 10,972 & 415,912 \\
        \hline
        r & Office~\cite{Genois2015DataLinkers} & Contacts between individuals in an office building & Contact & Spatial & 92 & 9,827 \\
        \hline
        s & AskUbuntu~\cite{Paranjape2017MotifsNetworks} & User answers or comments on questions on AskUbuntu & Social & Clique & 159,316 & 964,437 \\
        \hline
        t & MathOverflow~\cite{Paranjape2017MotifsNetworks} & User answers or comments on questions on MathOverflow  & Social & Clique & 24,818 & 506,550 \\
        \hline
        u & StackOverflow~\cite{Paranjape2017MotifsNetworks} & User answers or comments on questions on StackOverflow  & Social & Clique & 2,601,977 & 63,497,050 \\
        \hline
        v & SuperUser~\cite{Paranjape2017MotifsNetworks} & User answers or comments on questions on SuperUser  & Social & Clique & 194,085 & 1,443,339 \\
        \hline
        x & UCLA AS~\cite{Clegg2009} & UCLA AS level internet topology & Computer & Individual & 38,219 & 226,729 \\
        \hline
        y & US Air Traffic~\cite{Paranjape2017MotifsNetworks} & Flights among all commercial airports in the US & Transport & Individual & 2,278 & 6,390,340 \\
        \hline
        z & Wiki Talk~\cite{Paranjape2017MotifsNetworks} & Wikipedia users editing each other's Talk page  & Social & Individual & 1,140,149 & 7,833,140 \\
        \hline
        A & IETF~\cite{khare2022untangling} & IETF mailing list replies & Social & Clique & 23,792 & 989,740 \\
    \end{tabular}}
        \caption{Dataset corpus used for network creation.}
    \label{tab:datasets}
\end{table}

For each network, we treat each edge event as one time iteration for the network, where all edges that happen simultaneously are grouped into the same edge event. The whole time period is split into five equal time windows, this is to ensure any long periods of inactivity in the network do not leave empty spaces in our plots. Then, we focus on the first window and divide it into two windows of equal length. We build a network snapshot for both of these two sub-windows, and calculate the mobility taxonomy correlations between them. Then, we push the time window forward by $\frac{1}{2}$ of its duration. This is repeated until the time window reaches the end of the whole time period of the data. Therefore, we calculate 9 correlations for each network, each taking into account $\frac{1}{5}$ of the total time period of the network.

\begin{figure}[!tb]
\centering{\includegraphics[width=1.0\textwidth]{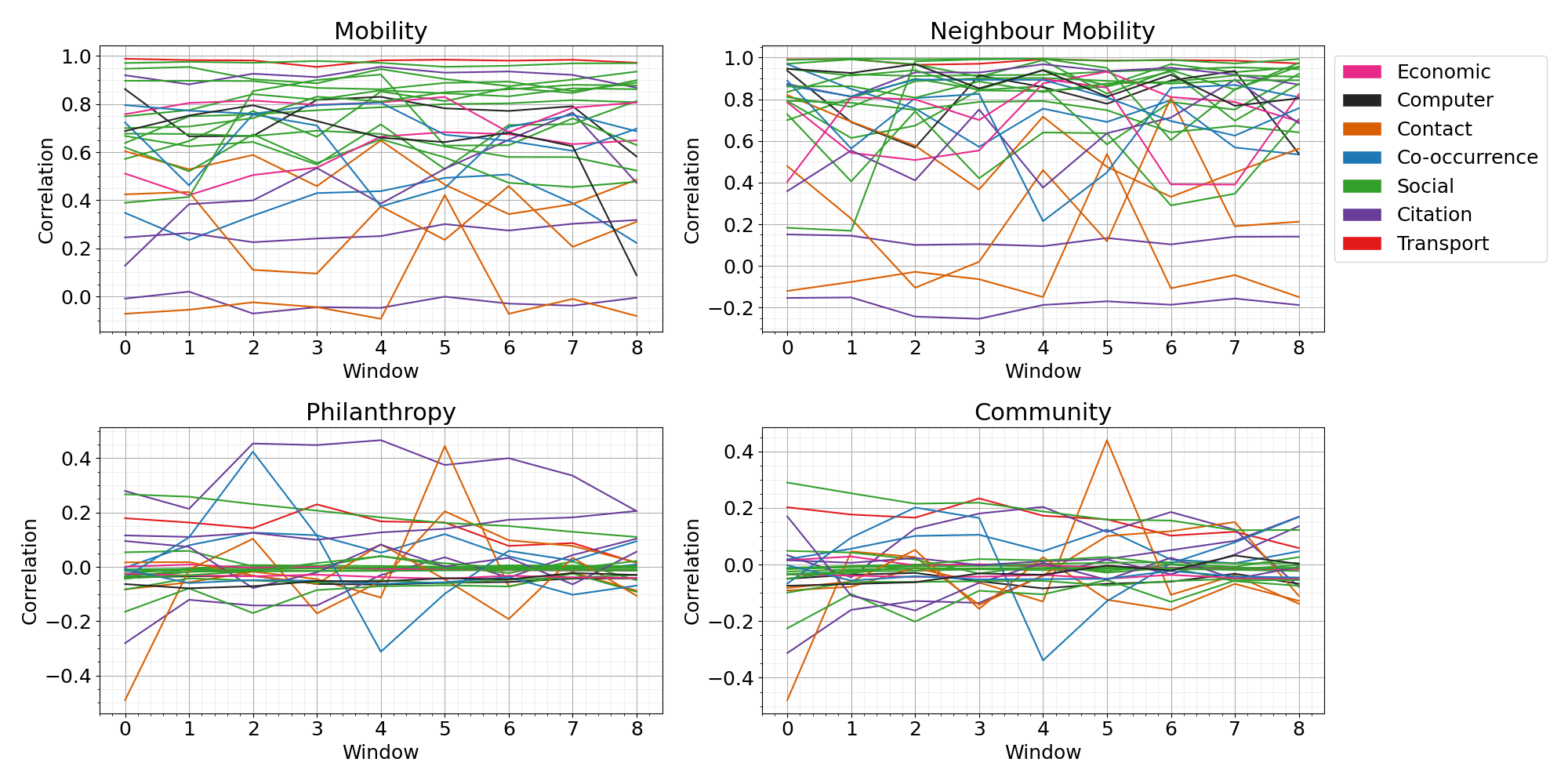}
  \caption{All networks' mobility taxonomy correlations in the corpus coloured by data type. The x-axis represents the ordered window numbers starting at the first timestamp of the network. The y-axis shows the correlation coefficient as measured using Pearson correlation.}
\label{fig:taxonomy_lines}
}
\end{figure}

In Figure~S\ref{fig:taxonomy_lines}, we plot each network in the corpus individually. This plot is much harder to interpret than Fig. 1 which uses standard deviations to give insight into the ranges of correlation values for each network collection type. However, this plot clearly shows the two low {\it mobility} and {\it neighbour mobility} networks which are referred to in the main paper. The first of these is Citation US Patents~\cite{Hall2001TheTools} which has unusually high {\it philanthropy} correlation suggesting patents citing highly cited patents themselves receive a large amount of patents. The other is Infectious~\cite{Isella2011WhatsNetworks} which has a recurring ``jump'' in each correlation, this is due to a sudden large decrease in the amount of active nodes in the network (the authors have not managed to verify the cause of this decrease). Finally, the large shift in the Lord Of the Rings~\cite{Clegg2009} philanthropy and {\it community} correlations, as referred to in the main paper, can be seen more easily in this plot.

\section{Extra artificial data results}

Table~\ref{tab:models} describes the twelve models used in the artificial data study. The models chosen were taken from the literature combined with models that the authors reasoned might maximise or minimise the taxonomy statistics being investigated. As mentioned in the main text, each model uses the basic structure of the classic preferential attachment model~\cite{Barabasi1999EmergenceNetworks} where one node is introduced at every iteration and three links. The three links are connected to an existing node with probabilities chosen according to certain rules, for example the classic preferential attachment rule is that the probability of connecting to a node is precisely proportional to its degree.

\begin{table}[!tbp]
    \resizebox{\textwidth}{!}{\begin{tabular}{c|c}
        Name & Probability of Receiving Edge \\
         \hline \hline
        Random & Each node is equal \\
        \hline
        Preferential attachment & Proportional to each node's individual degree \\
        \hline
        Preferential neighbour attachment & First, a node is chosen using preferential attachment (PA); then one of its neighbours is chosen using PA \\
        \hline
        Equality & Inversely proportional to each node's individual degree\\
        \hline
        Sum neighbour degree  & Proportional to the sum of the degree of the neighbours of each node \\
        \hline
        Average neighbour degree & Proportional to the average of the degree of the neighbours of each node \\
        \hline
        Inverse average neighbour degree & Inversely proportional to the average of the degree of the neighbours of each node \\
        \hline
        Cluster & Proportional to the clustering coefficient of each node \\
        \hline
        Eigen & Proportional to the eigenvector centrality of each node \\
        \hline 
        Fitness & Proportional to a sample taken from a $\Gamma(x)$ distribution \\
        \hline
        Gamma & Same as above but every node is re-sampled every $m$ iterations \\
        \hline
        Gamma individual & Same as above but only one node is re-sampled \\
    \end{tabular}}
        \caption{Models used for generating the evolutionary dynamics of artificial networks.}
    \label{tab:models}
\end{table}

\begin{figure}[!tb]
\centering{\includegraphics[width=0.8\textwidth]{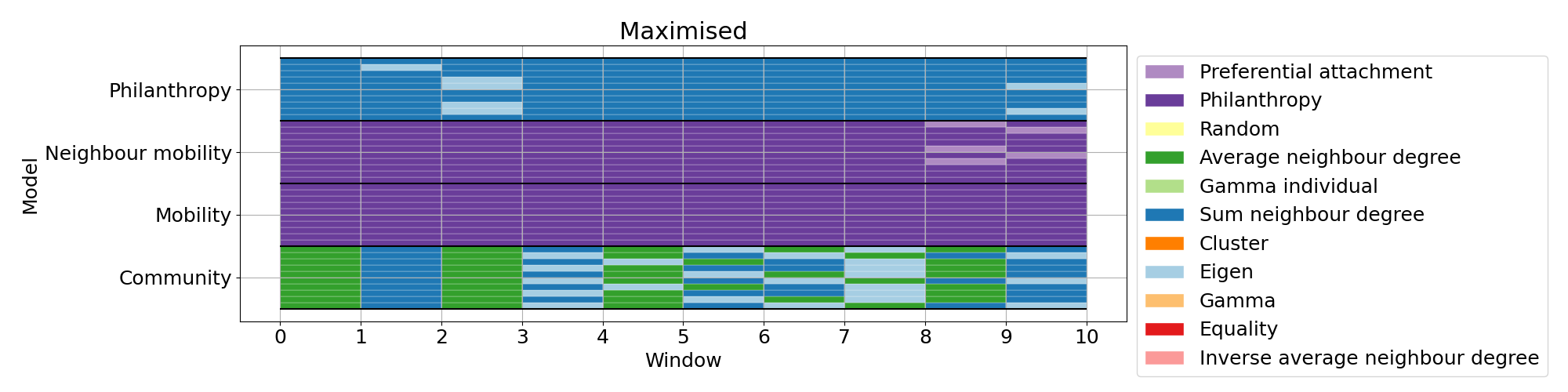}
  \caption{Visualisation of the evolution models used in the maximisation of mobility taxonomy aspects. The x-axis represents the number of 1,000 iteration-long windows (not including the initialisation of the graph).}
\label{fig:optimiser_runs_max}
}
\end{figure}

\begin{figure}[!tb]
\centering{\includegraphics[width=0.8\textwidth]{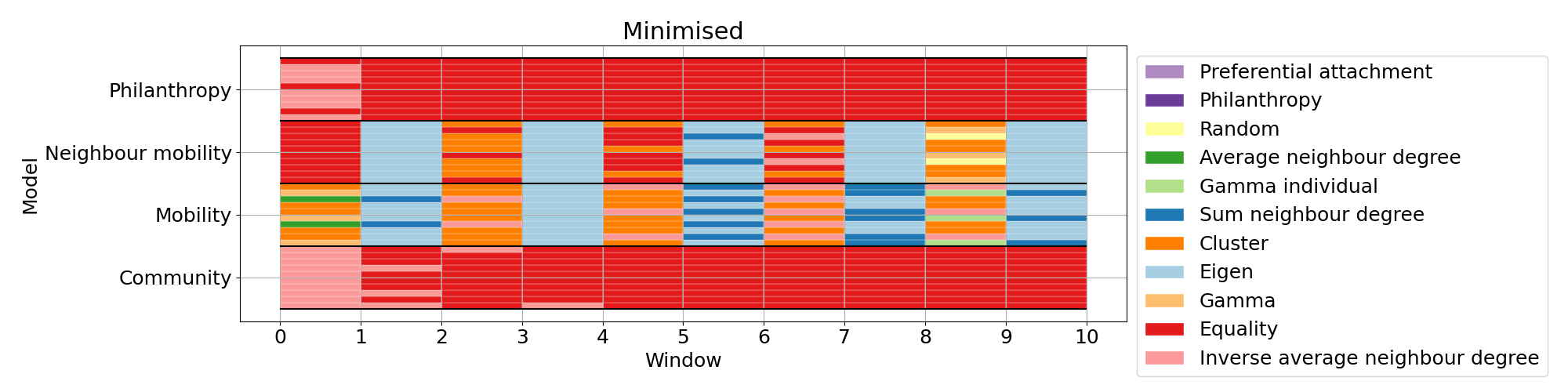}
  \caption{Visualisation of the evolution models used in the minimisation of mobility taxonomy aspects. The x-axis represents the number of 1,000 iteration-long windows (not including the initialisation of the graph).}
\label{fig:optimiser_runs_min}
}
\end{figure}

The figures~\ref{fig:optimiser_runs_max} and ~\ref{fig:optimiser_runs_min} show precisely which models were used in the maximisation and minimisation experiments. Each of the ten runs for each experiment is a separate row. For maximisation we can see that for maximising neighbour mobility and mobility the strategy was almost always philanthropy with occasionally the preferential attachment strategy. It makes sense that the strategy that maximises the correlation between degree in one time slice and the next does the same for a neighbourhood. For maximising philanthropy the sum neighbour degree was usually the most successful with occasionally eigen. Finally maximising community took an alternating strategy with average neighbour degree alternating with eigen and sum neighbour degree.

For minimisation equality and sometimes inverse average neighbour degree minimised philanthropy and community. These two strategies would tend to produce networks where all nodes have a similar degree. Minimising neighbour mobility and mobility itself both took an alternating strategy with a complex mix of strategies involved. 

\printbibliography